\documentclass[twocolumn,prl,aps,floatfix,superscriptaddress,amssymb]{revtex4}
\usepackage{epsfig}
\usepackage{amsmath}

\begin{document}

\title{Excitonic absorption in gate controlled graphene quantum dots}

\author{A.~D.~G\"u\c{c}l\"u}
\affiliation{Institute for Microstructural Sciences, National Research Council of Canada
, Ottawa, Canada}

\author{P.~Potasz}
\affiliation{Institute for Microstructural Sciences, National Research Council of Canada
, Ottawa, Canada}
\affiliation{Institute of Physics, Wroclaw University of Technology, Wroclaw, Poland}

\author{P.~Hawrylak}
\affiliation{Institute for Microstructural Sciences, National Research Council of Canada
, Ottawa, Canada}

\date{\today}

\begin{abstract}
We present a theory  of excitonic processes in gate controlled graphene
quantum dots.  The dependence of the energy gap on shape, size and edge for
graphene quantum dots with up to a million atoms is predicted. Using a
combination of tight-binding, Hartree-Fock and configuration interaction
methods, we show that triangular graphene quantum dots with zigzag edges
exhibit  optical transitions simultaneously in the THz, visible and UV
spectral ranges, determined by strong electron-electron and excitonic
interactions. The relationship between optical properties and finite magnetic
moment and charge density controlled by an external gate is predicted.
\end{abstract}

\maketitle

Two-dimensional graphene monolayer exhibits fascinating electronic
\cite{Wallace,NGM+04,NGM+05,ZTS+05,ZGG+06,NGP+09} and optical
properties\cite{Blinowski,Hoffman,SMP+06,NBG+08,WZT+08,LHJ+08,MSW+08,MXA10,YDP+09}
due to the zero energy gap and relativistic-like nature of quasiparticle
dispersion close to the Fermi level. With recent improvements in
nanofabrication techniques\cite{CMS+09p} the zero energy gap of bulk graphene
can be opened via engineering size, shape, character of the edge and carrier
density, and this in turn offers possibilities to simultaneously control
electronic\cite{BYB+05,IGG+09,PSK+08,WSG08,WRA+09,LSB09,AHM08,Eza10,PGH10},
magnetic\cite{CMS+09p,AHM08,Eza10,PGH10,GPV+09,Eza07,FP07,WMK08,YNW06} and
optical\cite{YNW06,ZCP08,YCL+10} properties of a single-material
nanostructure. 

In this paper, we present a theory and results of numerical calculations
predicting the dependence of the energy gap on shape, size and edge for
graphene quantum dots with up to a million atoms. We show that triangular
graphene quantum dots with zigzag edges combine magnetic and optical
properties tunable with carrier density, with optical transitions
simultaneously in the THz, visible and UV spectral ranges.  
We describe one electron properties of graphene quantum dots with $N$ atoms
and $N_e$ $\pi_z$ electrons by   a combination of tight-binding  approach with
a self-consistent Hartree-Fock method (TB-HF) described in detail in
the EPAPS document of our earlier work Ref.(\onlinecite{GPV+09}).  Then, in
order to take into account correlation and excitonic effects, we solve  the
many-body Hamiltonian given by
\begin{eqnarray}
&H&= \sum_{p',\sigma}\epsilon_{p'}b^\dagger_{p'\sigma}b_{p'\sigma}
    +\sum_{p,\sigma}\epsilon_{p}h^\dagger_{p\sigma}h_{p\sigma} \\
\nonumber
    &+&\frac{1}{2}\sum_{\substack{p'q'r's' \\ \sigma \sigma'}}\langle p'q'\vert V_{ee} \vert r's'\rangle 
     b^\dagger_{p'\sigma}b^\dagger_{q'\sigma'}b_{r'\sigma'}b_{s'\sigma}\\
\nonumber
    &+&\frac{1}{2}\sum_{\substack{pqrs \\ \sigma \sigma'}}\langle pq\vert V_{ee} \vert rs\rangle
     h^\dagger_{p'\sigma}h^\dagger_{q'\sigma'}h_{r'\sigma'}h_{s'\sigma}\\
\nonumber
    &+&\sum_{\substack{p'qrs' \\ \sigma \sigma'}}
     (\langle rp'\vert V_{ee} \vert s'q\rangle-\delta_{\sigma \bar{\sigma}'}
      \langle rp'\vert V_{ee} \vert qs'\rangle)
     b^\dagger_{p'\sigma}h^\dagger_{q\sigma'}h_{r\sigma'}b_{s'\sigma}\\
\nonumber
\label{CIhamilton}
\end{eqnarray}
where $b^\dagger_{p'\sigma}$ and $h^\dagger_{p\sigma}$ are hole and  electron
creation operators corresponding to TB-HF quasi-particles. Excitonic
absorption spectrum between ground state $\vert \nu_G \rangle$ and excited
states $\vert \nu_f \rangle$  can then be calculated using
\begin{eqnarray}
A(\omega)=\sum_{f} \vert \langle \nu_G \vert P \vert \nu_f \rangle \vert
          \delta(\omega-[E_f-E_G]))
\end{eqnarray} 
where $P=\sum_{pp'}\delta_{\sigma \bar{\sigma}'} \langle p \vert {\bf r} \vert
p' \rangle h_{p\sigma}b_{p'\sigma'}$ is the polarization operator.

\begin{figure}
\epsfig{file=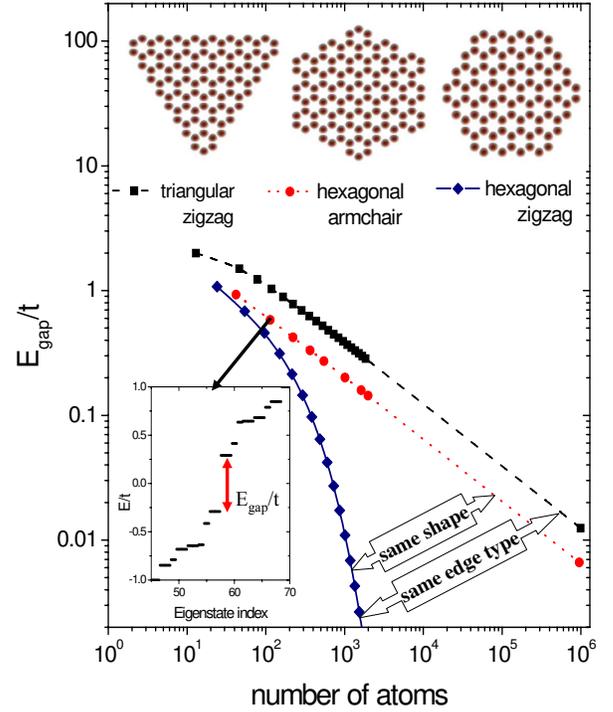,width=3.2in}
\caption{\label{fig:gap} (Color online)  Tight-binding bandgap energy as a
  function of total number of atoms N for a triangular zigzag quantum dot
  (black squares), hexagonal armchair quantum dot (red circles), and hexagonal
  zigzag quantum dot (blue diamonds). The inset shows the tight-binding energy
  spectrum in the vicinity of the Fermi level for the hexagonal armchair dot.}
\end{figure}

The electronic properties of graphene quantum dots depend on the size, shape
and the character of the edge. This is illustrated by comparing electronic
properties of three graphene quantum dots including (i) hexagonal dot with
armchair edges, (ii) hexagonal dot with zigzag edges, and (iii) triangular dot
with zigzag edges (see the inset of Fig. 1). The electronic structures are
computed using tight-binding Hamiltonian only with nearest-neighbor hopping
for different number of atoms $N$. An example of the energy levels for $N=98$
hexagonal quantum dot with armchair edges is shown in the inset of Fig. 1. The
red arrow indicates the bandgap separating the occupied valence band states
from the empty conduction band states. The dependence of the gap on the number
of atoms is plotted in Fig.1. For the hexagonal dot (red circles), the gap
decays as the inverse of the square root of number of atoms N, from hundred to
million atom nanostructures. This is expected for confined Dirac fermions with
photon-like linear energy dispersion ($E_{gap}\propto k_{min} \approx 2\pi /
\Delta x \propto 1/ \sqrt{N}$), as pointed out in
Refs.(\onlinecite{ZCP08,AHM08,RL09}). However, the replacement of the edge
from hexagonal to the zigzag has a significant effect on the energy gap. The
energy gap of hexagonal structure with zigzag edges decreases rapidly as the
number of atoms increases. This is due to the zigzag edges leading to
localized states at the edge of the quantum dot, similar to whispering gallery
modes of photons localized at the edge of photonic microdisk\cite{AKS+03}.
Fig. 1 also shows the effect on the energy gap of deforming the hexagonal
structure into a triangle while keeping zigzag edges. The energy gap of a
triangular dot is extracted from the energy spectrum, an example of which is
shown in Fig. 2(a) for a triangular quantum dot with N=96. In addition to
valence and conduction bands, the spectrum shows a shell of degenerate levels
at the Fermi level \cite{AHM08,PGH10,GPV+09,Eza07,FP07,WMK08,YNW06}. The
energy gap shown in Fig. 1 corresponds to transitions from the topmost valence
to the lowest conduction band state (see Fig. 2(a). Despite the presence of
the zero energy shell, the energy gap in the triangular zigzag structure
follows the power law $E_{gap}\propto \sqrt{N}$. We note that the energy gap
changes from $\approx 2.5$ eV (green light) for a quantum dot with $\approx
100$ atoms to $\approx 30$ meV (8 THz) for a quantum dot with a million atoms
and a diameter of $\approx 100$ nm. The presence of a partially occupied band
of degenerate states in the middle of a well defined energy gap offers unique
opportunity to simultaneously control magnetic and optical properties of
triangular graphene nanostructures. Indeed, due to the presence of the
zero-energy band in the middle of the energy gap, several different photon
energies corresponding to transitions within the zero-energy band, into and
out of the zero-energy states, and valence-to-conduction band states are
possible. This offers interesting possibilities for opto-electronic,
opto-spintronic and intermediate-band solar cell photo-voltaic applications
\cite{YCL+10,LM10}.

\begin{figure}
\epsfig{file=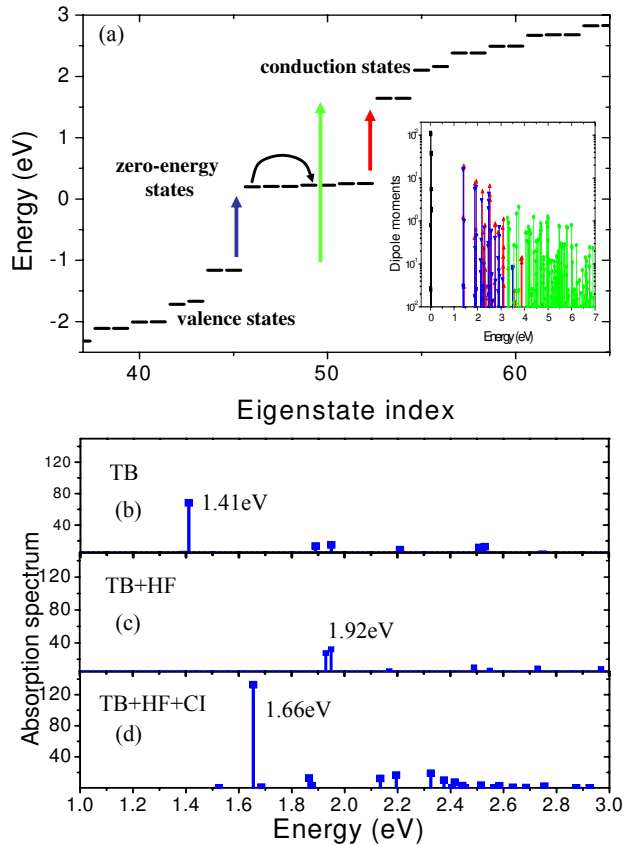,width=3.3in}
\caption{\label{fig:absorpTB} (Color online)   (a) Single particle
  tight-binding energies of states near Fermi level for a N=96-atom triangular
  zigzag quantum dot. The colored arrows represent optical transitions from
  valence to conduction band (VC, green), valence to zero-energy band (VZ,
  blue), zero-energy to conduction band (ZC, red) and zero-energy to
  zero-energy band (ZZ, black). Inset shows corresponding dipole moments for
  all transitions up to 7 eV, obtained using tight-binding
  orbitals. Figs. (b-d) shows the effect of electron-electron interactions on
  the VZ transitions within (c) Hartree-Fock approximation, and including
  (d) correlations and excitonic effect obtained from exact configuration
  interaction calculations.}
\end{figure}

The dependence of optical and magnetic properties on the filling of the shell
of degenerate zero-energy states is investigated in detail for a small
triangular dot with zigzag  edges containing N=96 atoms. For this system size,
exact many-body calculations can be carried out. The degenerate band in the
energy spectrum, shown in Fig. 2(a), has $N_z=7$ zero-energy states. Each
state is singly occupied and all electrons have parallel spin
\cite{GPV+09,Eza07,FP07,WMK08}. We can thus classify allowed optical
transitions into four classes, as shown in Fig. 2(a): (i) from valence band to
zero-energy degenerate band (VZ transitions, blue color); (ii) from
zero-energy band to conduction band (ZC transitions, red color); (iii) from
valence band to conduction band (VC transitions, green color); and finally,
(iv) within zero-energy states (ZZ transitions, black color). As a
consequence, there are three different photon energy scales involved in the
absorption spectrum. The corresponding joint optical density of states,
calculated using dipole moments $\vert \langle i \vert {\bf r} \vert f \rangle
\vert^2$ connecting initial and final states with energies $E_i$’s and
$E_f$’s, are shown in the inset of Fig. 2(a). VC transitions (green) occur
above full bandgap ($\approx 2.8$ eV), VZ (blue) and ZC (red) transitions
occur starting at half bandgap ($\approx 1.4$ eV), and ZZ (black) transitions
occur at THz energies. The energies corresponding to ZZ transitions are
controlled by the second nearest-neighbor tunneling matrix element $t_2$ and
by electron-electron interactions. 

Fig. 2(b-d) illustrates in detail the effect of electron-electron and
final-state (excitonic) interactions on the absorption spectra. Fig. 2(b)
shows detailed VZ absorption spectrum for noninteracting electrons. This
spectrum corresponds to transitions from the filled valence band to half
filled shell of $N_z=7$ zero-energy states. Half-filling implies that each
state of the zero-energy band is optically allowed. Numerical and analytical
calculations show that among the $N_z=7$ zero-energy states there are two bulk
like states, which couple strongly to the valence band resulting in the main
transition at $E=1.41$ eV. The remaining peaks in the absorption spectrum
reflect transitions from the discrete spectrum of the valence band. When the
electron-electron interactions are turned on, the ground state becomes a
Slater determinant of self-consistently calculated Hartree-Fock orbitals:
Valence states are doubly occupied, and zero-energy states are singly occupied
by an electron with the same spin, e.g., down. The photon energies
corresponding to optical transitions $\omega=(E_f+\Sigma_f)-(E_i+\Sigma_i)$
are renormalized by the difference in quasi-particle self-energies
$\Sigma_f-\Sigma_i$. The absorption spectrum, shown in Fig. 2(c), is
renormalized with transition energies blue-shifted by $0.51$ eV to $E=1.92$
eV. Finally, when final state interactions between all interacting
quasi-electron and quasi-hole states are taken into account, the excitonic
spectrum is again renormalized from the quasi-particle spectrum, with
transitions red shifted from quasi-particle transitions at $E=1.92$ eV, down
to $E=1.66$ eV. As we can see, electron-electron interactions play an
important role in determining energies and form of the absorption spectrum,
with net blue shift from the noninteracting spectrum by $0.25$ eV.

\begin{figure}
\epsfig{file=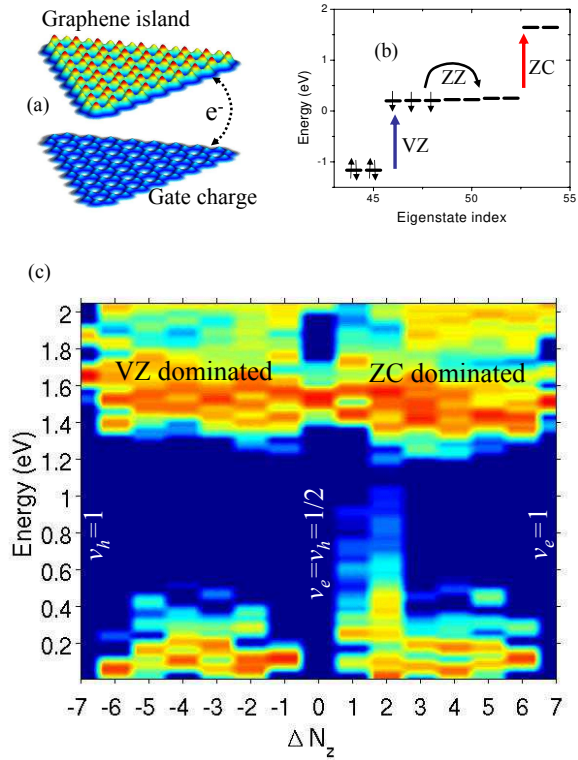,width=3.2in}
\caption{\label{fig:absorpX} (Color online)  (a) Schematic representation of a
  TGQD with $N=97$ carbon atoms with 4 electrons moved to the metallic
  gate. (b) Corresponding single particle tight-binding configuration near the
  Fermi level. (c) Excitonic absorption spectrum as a function of energy and
  charging $\Delta N_z$. For convenience, transitions are artificially
  broadened by 0.02 eV. Peaks below 0.6 eV are due to ZZ transitions, peaks
  above $1.2$ eV are due to VZ (mostly on the left), and ZC (mostly on the
  right) transitions. Charge neutral case corresponds to $\Delta N_z=0$
  (filling factors $\nu_e=\nu_h=1/2$). At $\Delta N_z=7$, the degenerate band
  of zero states is completely filled with electrons ($\nu_e=1$).}
\end{figure}

We now turn to the analysis of the effect of carrier density on the optical
properties of graphene quantum dots. The finite carrier density, controlled by
either metallic gate or via doping (intercalation), has been shown to
significantly modify optical properties of graphene
\cite{Blinowski,Hoffman,WZT+08,LHJ+08}.  For a quantum dot, the metallic gate
shown in Fig. 3(a), changes the number of electrons in the degenerate shell
from $N_z$ to $N_z+\Delta N_z$. This is illustrated in Fig. 3(b) where 4
electrons were removed and 3 electrons remain. These remaining electrons
populate degenerate shell and their properties are entirely controlled by
their interaction. Alternatively, removal of electrons from charge neutral
shell corresponds to addition of “holes”. As is clear from Fig. 3(b), such a
removal of electrons allows intra-shell transitions ZZ, enhances VZ
transitions by increasing the number of allowed final states and weakens the
ZC transitions by decreasing the number of occupied initial states. Fig. 3(c)
illustrates the overall effects in the computed excitonic absorption spectra
for VZ, ZC, and ZZ transitions as a function of the number of additional
electrons $\Delta N_z$, related to shell filling and/or gate voltage. At
$\Delta N_z=-7$ (hole filling factor $\nu_h=1$), the shell is empty and VZ
transitions describe an exciton built of a hole in the valence band and an
electron in the degenerate shell. The absorption spectrum has been described
in Fig. 2(b-d) and is composed of one main excitonic peak at 1.66 eV. There
are no ZC transitions and no ZZ transitions in the THz range. When we populate
the shell with electrons, the VZ excitonic transition turns into a band of red
shifted transitions corresponding to an exciton interacting with additional
electrons, in analogy to optical processes in the fractional quantum Hall
effect and charged semiconductor quantum dots\cite{HK03}. As the shell filling
increases, the number of available states decreases and the VZ transitions are
quenched while ZC and ZZ transitions are enhanced. These results show that the
absorption spectrum can be tuned by shell filling, which can be experimentally
controlled by applying a gate voltage. This is particularly true for the ZZ
transitions in the THz range, which can be turned off by either
emptying/filling the shell $\Delta N_z=\pm 7$ or at half-filling. At
half-filling, electron exchange leads to spin polarization, with each state of
the shell filled by a spin polarized electron. Since photons do not flip
electron spin, no intra-shell transitions are allowed and the magnetic moment
of graphene quantum dot is directly reflected in the ZZ absorption
spectrum. In addition, it was shown that even though at charge neutrality
($\Delta N_z=0$) the system is fully spin polarized, addition (but not
subtraction) of a single electron leads to full spin
depolarization\cite{GPV+09}.  This correlation effect plays an important role
in the optical transitions involving zero energy states, shown in Fig. 3(c)
for $\Delta N_z=\pm 1$.

\begin{figure}
\epsfig{file=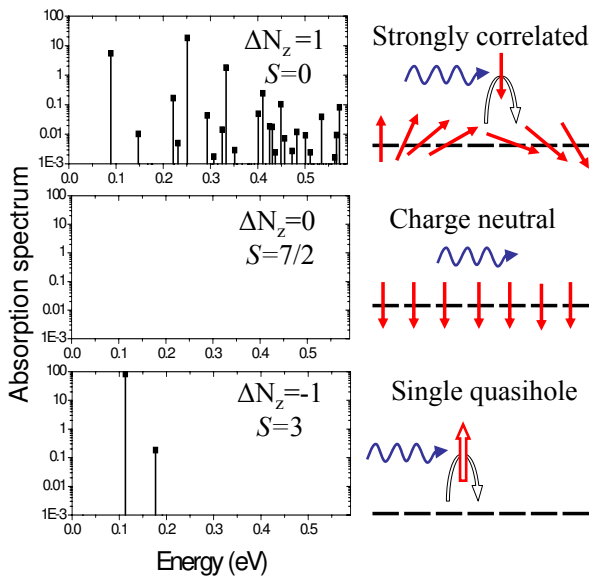,width=3.2in}
\caption{\label{fig:absorpX2} (Color online)  Excitonic absorption spectrum at
  $\Delta N_z=$-1, 0, and 1 (on the left). Corresponding ground state spins
  are $S=3$ (fully polarized), $S=7/2$ (fully polarized), and $S=0$
  (completely depolarized), respectively. The physics involved in optical
  transitions is schematically represented on the right side.  }
\end{figure}

In Fig. 4, we study the transitions for $\Delta N_z=0,\pm 1$ in
detail. Fig. 4(b) shows the lack of absorption for half-filled spin polarized
shell. The right hand side illustrates the fact that photons pass through
since they are not able to induce electronic transitions and be absorbed. For
$\Delta N_z=-1$, Fig. 4(c), one electron is removed creating a hole in the
spin polarized shell. Thus, the absorption spectrum corresponds to transitions
from ground state to optically allowed excited states of the hole. The
absorption spectrum for an additional electron, $\Delta N_z=+1$, shown in
Fig. 4(a), is dramatically different. The additional electron depolarizes the
spins of all electrons present, with total spin of the ground state
$S=0$\cite{GPV+09}. The strongly correlated ground state has many
configurations, which effectively allow for many transitions of the spin up
and spin down electrons, as seen in Fig. 4(a). This asymmetry in the THz
absorption spectra allows for the optical detection of charge of the quantum
dot and correlated electron states in the degenerate electronic shell.


{\it Acknowledgement}. The authors thank M. Korkusinski, O. Voznyy, A. Wojs
and M. Potemski for discussions and NRC-CNRS CRP, Canadian Institute for
Advanced Research, Institute for Microstructural Sciences, and QuantumWorks
for support.

\vspace*{-0.22in}


\end{document}